\documentclass[12pt,english,floatfix,superscriptaddress,aps,prd,preprint]{revtex4}
\usepackage[utf8]{inputenc}
\usepackage{amsmath}
\usepackage{amssymb}
\usepackage{amsbsy}
\usepackage{amsfonts}
\usepackage{amsopn}
\usepackage{amstext}
\usepackage{graphicx}
\usepackage{amssymb}
\usepackage{amsfonts}
\usepackage{amsmath}
\usepackage{amsmath,amsthm,amsfonts,amssymb}
\usepackage[mathcal]{eucal}
\usepackage{mathrsfs}
\usepackage{graphicx}
\usepackage[english]{babel}
\usepackage{color}
\usepackage{esint}
\usepackage[dvips]{epsfig}
\usepackage[dvips]{graphicx}
\usepackage{float}
\usepackage{units}
\usepackage{textcomp}

\usepackage{hyperref}
\usepackage{slashed}

\newcommand{\ie}{\begin{equation}}
\newcommand{\fe}{\end{equation}}
\newcommand{\se}{\begin{eqnarray}}
\newcommand{\ff}{\end{eqnarray}}

\usepackage{xcolor}
\usepackage{soul}

\begin{document}

\title{ Traversable wormhole solution with a background Kalb-Ramond field }


\author{L. A. Lessa}
\email{leandrolessa@fisica.ufc.br}
\affiliation{Universidade Federal do Cear\'a (UFC), Departamento de F\'isica,\\ Campus do Pici, Fortaleza - CE, C.P. 6030, 60455-760 - Brazil.}
\author{R. Oliveira}
\email{rondinelly@fisica.ufc.br}
\affiliation{Universidade Federal do Cear\'a (UFC), Departamento de F\'isica,\\ Campus do Pici, Fortaleza - CE, C.P. 6030, 60455-760 - Brazil.}

\author{J. E. G. Silva}
\email{euclides.silva@ufca.edu.br}
\affiliation{Universidade Federal do Cariri (UFCA), Av. Tenente Raimundo Rocha, \\ Cidade Universit\'{a}ria, Juazeiro do Norte, Cear\'{a}, CEP 63048-080, Brazil}


\author{C. A. S. Almeida}
\email{carlos@fisica.ufc.br}
\affiliation{Universidade Federal do Cear\'a (UFC), Departamento de F\'isica,\\ Campus do Pici, Fortaleza - CE, C.P. 6030, 60455-760 - Brazil.}

\date{\today}

\begin{abstract}
We obtain a static spherically symmetric wormhole solution due to the vacuum expectation value (VEV) of a Kalb-Ramond field. The Kalb-Ramond VEV is a background tensor field which produces a local Lorentz symmetry breaking (LSB) of spacetime. Considering a non-minimal coupling between the Kalb-Ramond (VEV) and the Ricci tensor, we found an exact traversable wormhole solution sustained by matter sources with a negative isotropic pressure. The matter source satisfies the energy conditions at the throat for particular choices of the LSB parameter. Moreover, we employ the Gauss-Bonnet method to analyze the deflection angle of light in the weak limit approximation. 

\end{abstract}

\maketitle

\section{Introduction}
Wormholes are interesting solutions of Einstein equations whereby two distinct regions of space-time are connected by a throat region, thereby establishing a "shortcut" between these regions \cite{Fla}.
In 1935, a wormhole solution was proposed by Einstein and Rosen, known as the Einstein-Rosen bridge \cite{ER}. However, this solution proved to be very unstable, not allowing even photons to pass through it. It was only in 1988 that Morris and Thorne found a stable wormhole solution \cite{MT}. This solution requires an exotic matter source with negative energy density \cite{MT}. Consequently, the null energy condition $NEC$ is violated \cite{Mauricio}. In more recent works, new traversable wormholes without the need for exotic matter were proposed \cite{Can, Har}. Indeed, some modified gravity theories, as the $f(T)$  gravity \cite{Lobo}, $ f(R, \phi)$ gravity \cite{Baha}, Einstein-Cartan gravity \cite{Bron}, Lovelock gravity \cite{Zan}, lead to modified equation of state (E0S) even for a normal matter source.

Modified gravitational dynamics also arise due to local Lorentz violation effects \cite{vsr,dsr,horava}. Although the Lorentz symmetry might be broken near the Planck scale, effective field theories can capture reminiscent effects in the IR scale \cite{kostelecky}. A spontaneous breaking of Lorentz symmetry occurs when one or more tensor fields acquire nonzero vacuum expectation values (VEV) \cite{kostelecky}. These background VEV tensors defines privileged directions in spacetime, thereby violating the particle local Lorentz symmetry \cite{kostelecky}. A self-interacting vector field, known as the bumblebee field \cite{ngmodes}, leads to modification of the gravitational propagation and interaction with matter \cite{Malufbumblebee}. Further, expanding cosmological solutions \cite{Capelo} and black hole solutions \cite{bertolami,casana,Ding,Ovgun2} were found driven by the Bumblebee field. In Ref.\cite{Jusu}, wormhole solutions were found where the VEV bumblebee vector enables a source satisfying the null and weak energy condition $WEC$. The wormhole turns out to satisfy the flare-out condition, i.e., without horizon \cite{Jusu,Rond}.   

Another source for the spontaneous Lorentz symmetry breaking (LSB) is the Kalb-Ramond (KR) field, an antisymmetric tensor field $B_{\mu\nu}$ which arises in string theories \cite{kalb}. Allowing a self-interaction potential, a non-vanishing VEV breakes the gauge and the Lorentz symmetry \cite{kalb2}. Likewise the electromagnetic field strength, the tensor antisimmetric VEV $b_{\mu\nu}$ can be decomposed into two vectors, known as pseudo-electric and pseudo-magnetic vectors \cite{kalb2}. Recently, a static and spherically symmetric black hole modified by a background KR field was found in Ref.\cite{leandro}. A non-minimal coupling between the KR VEV and the Ricci tensor lead to a parameter-dependent power-law correction to the Schwarszchild solution \cite{leandro}. The effects of this LSB solution was studied on the black hole temperature \cite{leandro} and the shadows \cite{Kumar}.

In this work, we propose a LSB wormhole solution driven by the Kalb-Ramond field. Likewise the bumblebee wormhole solution \cite{Jusu}, the non-minimal coupling of the KR field to the Ricci tensor allows a traversable wormhole with a matter source satisfying the null, weak and strong energy conditions for a range of the Lorentz violating parameter. In addition, the spacetime is asymptotically flat and the negative pressure is fined tuned by the Lorentz violation. By employing the Gauss-Bonnet theorem \cite{Gibbons,Kimet,Suzuki,Ju-Ov,Sakalli,Gun,Ara,Og, Ka}, we explored the effects of the LSB on the light deflection.

The work is organized as the following. In section (\ref{sec1}), we present a short review on the spontaneous symmetry breaking mechanism for the Lorentz symmetry driven by the Kalb-Ramond field. Moreover, we define the non-minimal coupling between the KR VEV and the Ricci tensor, and we choose the vacuum configuration. In section (\ref{sec2}) we obtain an exact wormhole solution for the modified Einstein equation found in section (\ref{sec1}). Moreover, we analyze the traversibility of the wormhole obtained. In section (\ref{sec3}), we check the energy conditions of the Kalb-Ramond wormhole. In section (\ref{sec4}) we study the deflection angle of light in the weak limit approximation. Finally, in section (\ref{sec5}), final comments and perspectives are outlined. Throughout the text, we adopt the metric signature $(-,+,+,+)$.


 \section{The Kalb-Ramond model for spontaneous Lorentz symmetry breaking} \label{sec1}

In this section, we present the Kalb-Ramond dynamics, discussing the Vaccum expectation value (VEV) choice and features and the corresponding coupling of this background field to gravity.

The Kalb-Ramond (KR) field is a rank two antisymmetric tensor field $B_{\mu\nu}$ which arises in the spectrum of bosonic string theory \cite{kalb}. In the differential form language, the KR field can be defined as a 2-form potential $B_2 = \frac{1}{2}B_{\mu\nu}dx^{\mu}\wedge dx^{\nu}$ whose field strength is given by $H_3 = dB_2$, i.e., $H_{\lambda\mu\nu} \equiv \partial_{[\lambda}B_{\mu\nu]}$. Likewise the photon field, the KR potential $B_2$ possesses a gauge symmetry, namely $B_2 \rightarrow B_2 + d\Lambda_1$, where $\Lambda_1$ is a 1-form.

By allowing a KR self-interaction some theories lead to KR configurations which break both the gauge and Lorentz symmetries \cite{kostelecky}. 
In the gravitational sector of the Standard Model Extension (SME), a self-interacting potential of form $V = V(B_{\mu\nu}B^{\mu\nu} \pm b_{\mu\nu}b^{\mu\nu})$ with a non-vanishing (VEV) $<B_{\mu\nu}> =b_{\mu\nu}$ defines a background tensor field and thus, the Lorentz symmetry is spontaneously broken by the Kalb-Ramond self-interaction \cite{kostelecky,kalb2}. It is worthwhile to mention that the dependence of the potential on $B_{\mu\nu}B^{\mu\nu}$ keeps the theory invariant under observer local Lorentz transformations. 

Consider an action for a self-interacting Kalb-Ramond field non-minimal coupled with gravity in the form \cite{kalb2}
\begin{align} \nonumber \label{38}
&\mathcal{S}^{nonmin}_{KR} =  \int e \ d^{4}x \bigg[ \frac{R}{2\kappa} - \frac{1}{12} H_{\lambda\mu\nu}H^{\lambda\mu\nu} - V(B_{\mu\nu}B^{\mu\nu} \pm b_{\mu\nu}b^{\mu\nu}) \\
&+\frac{1}{2\kappa} \bigg(\xi_{2}B^{\lambda\nu}B^{\mu} \ _{\nu}R_{\lambda\mu}+ \xi_{3}B^{\mu\nu}B_{\mu\nu}R \bigg)  +  \mathcal{L}^{M}  \bigg],
\end{align}
where $\xi_{2}$ and $\xi_{3}$ are non-minimal coupling constants (with dimensions [$\xi$]=$L^{2}$), $e$ is the metric determinant and $\kappa =8 \pi G $ is the gravitational coupling constant. The first and second terms are the usual Einstein-Hilbert and the KR field strength terms, whereas the third is the self-interaction potential responsible for the spontaneous breaking of the Lorentz symmetry.  It is worthwhile to mention that the non-minimal coupling enables a derivative interaction of the metric with the KR VEV which is at most of second order on the metric.

In this work we focus on the gravitational effects of a wormhole in a Lorentz symmetry broken phase, and thus, we consider the KR field in its vacuum configuration, i.e.,  $B_{\mu\nu}B^{\mu\nu}=b_{\mu\nu}b^{\mu\nu}$. In the absence of gravity, the flat spacetime structure enables a constant Lorentz violating VEV $b_{\mu\nu}$, e.g., $\partial_\rho b_{\mu\nu}=0$ \cite{kalb2}. Accordingly, the norm $b^2 = \eta^{\mu\nu}\eta^{\alpha\beta}b_{\mu\alpha}b_{\nu\beta}$ is constant and the corresponding Lorentz violating coefficients can be defined throughout the spacetime using $b_{\mu\nu}$ \cite{kalb2}. Moreover, a constant $b_{\mu\nu}$ yields to a vanishing KR field strength $h_3 =db_2$ and hence, to a vanishing VEV Hamiltonian \cite{kalb2}. A straightforward extension for curved spacetimes can is given by $\nabla_\rho b_{\mu\nu}=0$ which guarantees the vanishing of the the KR Hamiltonian vanish \cite{bertolami}. An alternative definition of the KR VEV is achieved assuming that $b_{\mu\nu}$ has constant norm $b^2=b_{\mu\nu}b^{\mu\nu}$, which is equivalent to $b^{\mu\nu}\nabla_\rho b_{\mu\nu}=0$ and ensures a vanishing potential \cite{casana,leandro}. Henceforward, we assume a constant norm KR VEV.

The antisymmetry of $b_{\mu\nu}$ allows us to decomposed it as 
$b_{\mu\nu}=\tilde{E}_{[\mu} v_{\nu]}+\epsilon_{\mu\nu\alpha\beta}v^\alpha \tilde{B}^{\beta}$, where the background vectors $\tilde{E}_\mu$ and 
$\tilde{B}_\mu$ can be interpreted as pseudo-electric and pseudo-magnetic fields, respectively, and $v^\mu$ is a timelike 4-vector \cite{leandro}. The pseudo-fields $\tilde{E}_\mu$ and $\tilde{B}_\mu$ are spacelike, i.e., $\tilde{E}_\mu v^\mu = \tilde{B}_\mu v^\mu=0$. Thus, the KB VEV yields two background vector instead of only one produced by the bumblebee VEV \cite{casana,bertolami}. 

Let us consider a pseudo-electric configuration of form
\begin{equation}
\label{vevansatz}
 b_{2} = - \tilde{E}(x^1) \ dx^{0} \wedge dx^{1}.
\end{equation}
The choice of the KR VEV in Eq.\eqref{vevansatz} enables us to write the VEV as $b_2 = d\tilde{A}_1$, where $\tilde{A}_1 = \tilde{A}_0 (x^1) dx^0$ can be interpreted as a pseudo-vector potential and $\tilde{E}=-\partial_1 A_0$. Such VEV configuration resembles a background electric field with vanishing Hamiltonian. The function $\tilde{E}(x^1)=-b_{01}$ will be determined by the condition $b^2$ constant in a wormhole geometry in the next section.

By varying Eq.(\ref{38}) with respect to the metric, the modified Einstein equations are \cite{kalb2,leandro}
\begin{equation} \label{39}
G_{\mu\nu}=\kappa T^{\xi_{2}}_{\mu\nu} + \kappa T^{M}_{\mu\nu},
\end{equation}
where $T^{M}_{\mu\nu}$ is matter field, with $G_{\mu\nu} = R_{\mu\nu} - \frac{1}{2}R g_{\mu\nu}$
and
\begin{align} \nonumber \label{43}
T^{\xi_{2}}_{\mu\nu} = & \dfrac{\xi_{2}}{\kappa} \bigg[ \dfrac{1}{2}g_{\mu\nu}B^{\alpha\gamma}B^{\beta} \ _{\gamma}R_{\alpha\beta} - B^{\alpha} \ _{\mu}B^{\beta} \ _{\nu}R_{\alpha\beta} \\ \nonumber
& - B^{\alpha\beta}B_{\mu\beta}R_{\nu\alpha} - B^{\alpha\beta}B_{\nu\beta}R_{\mu\alpha} \\ \nonumber
& + \dfrac{1}{2}D_{\alpha}D_{\mu}(B_{\nu\beta}B^{\alpha\beta}) + \dfrac{1}{2}D_{\alpha}D_{\nu}(B_{\mu\beta}B^{\alpha\beta}) \\
& - \dfrac{1}{2}D^{2}(B^{\alpha} \ _{\mu}B_{\alpha\nu}) - \dfrac{1}{2}g_{\mu\nu}D_{\alpha}D_{\beta}(B^{\alpha\gamma}B^{\beta} \ _{\gamma}) \bigg].
\end{align}
The absence of terms proportional to $\xi_{3}$ steams from the constancy of the VEV norm $b^2$, whereby the term $\xi_{3}B^{\mu\nu}B_{\mu\nu}R$ cab be absorbed into the usual Einstein-Hilbert Lagrangian by a redefinition of variables \cite{leandro}.
Therefore, the non-minimal coupling yields to a source modifying the field equation by new derivative terms. In the next section, we 
seek for wormhole solutions of Eq.\ref{39} endowed with a Lorentz violating KR VEV stress-energy given by Eq.(\ref{43}).


\section{Spherically symmetric Kalb-Ramond wormhole solutions}
\label{sec2}

Once we established the coupling of the background KR VEV to the gravitational field, let us look for a wormhole solution of modified Einstein equation Eq.(\ref{39}).

Consider a static and spherically symmetric traversable wormhole solution described by the metric \cite{MT},

\begin{equation}
\label{metric}
ds^{2} = -e^{2\phi(r)}dt^{2}+\frac{dr^{2}}{1-\frac{\Omega(r)}{r}} +r^{2}d\theta^{2}+r^{2}\sin^2\theta d\phi^{2},
\end{equation}
where $\phi(r)$ is the redshift function, presumably finite everywhere to
avoid the presence of event horizons \cite{MT}, and $\Omega(r)$ is shape function of the wormhole. 
Using the \textit{ansatz} metric in Eq.(\ref{metric}), the Kalb-Ramond VEV \textit{ansatz} given by Eq.\eqref{vevansatz} has a 
constant norm $b^2 =g^{\mu\alpha}g^{\nu\beta}b_{\mu\nu}b_{\alpha\beta}$ for
\begin{equation}
\label{vevdependence}
\tilde{E}(r) =  \frac{|b|e^{\phi(r)}}{\sqrt{2 \bigg( 1 - \frac{\Omega(r)}{r} \bigg)}},
\end{equation}
where $b$ is a constant. This background radial pseudo-electric static configuration $\tilde{E}^{\mu}=(0,\tilde{E},0,0)$ preserves the spheric and static spacetime symmetry, for the background vector $\tilde{E}^\mu$ is orthogonal to both the timelike $t^{\mu}=(\partial/\partial t)^\mu$ and spacelike $\psi^\mu =(\partial/\partial \phi)^\mu$ Killing vectors \cite{bertolami}.

Using the metric \textit{ansatz} \eqref{metric}, the modified Einstein equations \eqref{39} lead to system of equations

\begin{equation} \label{1}
G_{tt} = \frac{\lambda}{4} \bigg[ 3R_{tt} - \bigg(1-\frac{\Omega(r)}{r} \bigg)R_{rr} \bigg] + \kappa T^{M}_{tt} ,
\end{equation}

\begin{equation} \label{2}
G_{rr} = \frac{\lambda}{4} \bigg[ 3R_{rr} - \bigg(1-\frac{\Omega(r)}{r} \bigg)^{-1} R_{tt} \bigg] + \kappa T^{M}_{rr} ,
\end{equation}

\begin{equation} \label{3}
G_{\theta\theta} = \frac{\lambda r^{2}}{4}\bigg[ R_{tt} - \bigg(1-\frac{\Omega(r)}{r} \bigg)R_{rr} \bigg] + \kappa T^{M}_{\theta\theta} ,
\end{equation}

\begin{equation} \label{4}
G_{\phi\phi} = \sin^2\theta G_{\theta\theta},
\end{equation}
where $\lambda:= |b|^2\xi_{2}$.

From that point on, we neglect the effects of the tidal force from the Eq.(\ref{metric}), i.e, we assume that $\phi (r) = const$. Accordingly, the components of Ricci tensor are
\begin{equation}
R_{rr} = \frac{r \Omega'(r) - \Omega(r)}{r \bigg( 1 -\frac{\Omega(r)}{r}  \bigg)},
\end{equation}
\begin{equation}
R_{\theta\theta} = \frac{r \Omega'(r) + \Omega(r)}{2r},
\end{equation}
where the prime stands for the derivative with respect to $r$. $R_{tt}$ is zero due to the absence of the tidal force. 

In addition to the background KR field, we assume a matter stress-energy tensor in the form $(T^{\mu} \ _{\nu} )^{M}= (- \rho, p_{r},p_{\theta},p_{\phi})$. The energy density and the anisotropic pressures satisfy the modified Einstein's equations (\ref{1}), (\ref{2}), and (\ref{3}), provided that

\begin{equation} \label{1a}
\rho = \frac{1}{\kappa} \bigg[ \frac{\Omega'(r)}{r^{2}} + \frac{\lambda}{4r^{3}} \bigg(r\Omega'(r)-\Omega(r) \bigg)  \bigg],
\end{equation}

\begin{equation} \label{2a}
p_{r}= - \frac{1}{\kappa} \bigg[ \frac{\Omega(r)}{r^{3}} + \frac{3\lambda}{4r^{3}} \bigg(r\Omega'(r)-\Omega(r) \bigg),
\end{equation}

\begin{equation} \label{3a}
p_{\theta} = - \frac{1}{\kappa} \bigg( 1 - \frac{\lambda}{2} \bigg) \frac{r\Omega'(r)-\Omega(r)}{2r^{3}},
\end{equation}
\begin{equation}
p_{\phi} = p_{\theta}.
\end{equation}
A noteworthy feature of the energy density Eq.\eqref{1a} and the anisotropic pressures (\ref{2a}), and (\ref{3a}), is the absence of the
 terms $\Omega ''$ and $(\Omega ')^{2}$. This fact steams from the form of the wormhole metric \eqref{metric} and of the KR vev (\ref{vevdependence}) whose contributions cancel each other out, as also shown in Ref.\cite{Jusu}.

The task of achieve a feasible wormhole solution turns in to find $\rho$, $p_r$, $p_\theta$ and $\Omega$ from the  equations (\ref{1a}), (\ref{2a}), and (\ref{3a}). Assuming an isotropic ideal fluid, where $p=\alpha \rho$, the Einstein equations determines $\Omega$, $\rho$ and $\alpha$. Indeed, considering 
$p_{r}=p_{\theta}$ leads to $\Omega(r) \propto r^{\frac{3-2 \lambda}{1- 2 \lambda}}$ and requiring that $\Omega(r_{0})=r_{0}$, we find
the shape function as follows
\begin{equation} \label{shape}
\Omega(r) = r \bigg( \frac{r}{r_{0}}   \bigg) ^{\frac{2}{1-2 \lambda}}.
\end{equation}

Substituting the Eq. (\ref{shape}) in (\ref{1a}), the power-law shape function has a matter source whose energy density is
\begin{equation}
\label{energydensity}
 \rho = \bigg( 1 + \frac{2 + \frac{\lambda}{2}}{1- 2 \lambda}   \bigg) \frac{r^{\frac{4 \lambda}{1- 2 \lambda}}}{k r_{0} \ ^{\frac{2}{1- 2 \lambda}}}.
\end{equation}
Moreover, this ideal fluid satisfies the modified Einstein equations provided that 
\begin{equation}
\label{eos}
\alpha =  -\frac{1}{3}.
\end{equation}
Therefore, the source has a negative pressure tuned by the Lorentz violation. An example of source with equation of state Eq.(\ref{eos}) is furnished by a string gas.

 %

Once we found the shape function and the source, let us examine some geometric features. The spacetime metric takes the form
\begin{equation} \label{solucao2}
ds^2 = -dt^2 + \frac{dr^2}{1- \bigg( \frac{r}{r_{0}} \bigg)^{\frac{2}{1-2 \lambda}} } + r^2(d \theta ^2 + sen^2 \theta d \phi^2).
\end{equation}
Unlike the Ref. \cite{Mauricio}, we obtained a wormhole solution without tidal forces generated by a perfect fluid and a KR VEV. The Lorentz violation effects are encoded in the parameter $\lambda$. At the Lorentz invariant regime, i.e., $\lambda\rightarrow0$, for $|b|^2\rightarrow0$ or $\xi_2 \rightarrow 0$, we recover the zero-tidal-force wormholes solution with isotropic pressure, Ref.\cite{Mauricio}, as expected.

For $\lambda>0$ the Lorentz violation increases the power in the shape function by a factor of $(\frac{r}{r_0})^{\frac{1}{1-2\lambda}}$. If $\lambda\approx 0$, $\frac{1}{1-2\lambda}\approx 1+2\lambda$ and the asymptotic geometry is not flat, as in the Bumblebee Lorentz violating wormhole solution \cite{Jusu}. In order to obtain an asymptotically flat spacetime, the Lorentz violating parameter ought to be bound to the range $ \lambda > 1/2 $. The radial variable $r$
ranges from the minimal $r_0$ into infinity, i.e., $ r_{0} \leqslant r < \infty $, as required. A noteworthy feature of the KR Lorentz violating wormhole solution is the possibility to obtain asymptotically flat geometries, contrary to the Bumblebee field solution found in Ref.\cite{Jusu}.

From solution (\ref{solucao2}), the Ricci and Kretschmann scalars in $r=r_{0}$ are, respectively,
\begin{equation}
R|_{r_{0}} =2 \bigg(\frac{3-2 \lambda}{1- 2 \lambda}  \bigg) \frac{1}{r_{0}^{2}} ,
\end{equation}
\begin{eqnarray}  
K|_{r_{0}} =& \frac{4 \times 81^{\frac{1}{-1+2\lambda}} r_{0}^{\frac{8\lambda}{1-2\lambda}}}{(1-2\lambda)^{2}}  \bigg[1 + \bigg( \frac{81}{256}    \bigg)^{\frac{1}{1-2\lambda}}  + \bigg( \frac{81}{16}    \bigg)^{\frac{1}{1-2\lambda}} \nonumber \\ &-  2 \bigg(1 + \bigg( \frac{81}{256}    \bigg)^{\frac{1}{1-2\lambda}}  \bigg)\lambda \nonumber\\ 
&+  2 \bigg(1 + \bigg( \frac{81}{256}    \bigg)^{\frac{1}{1-2\lambda}}  \bigg)\lambda^2   \bigg].
\end{eqnarray}
For $\lambda\neq 0$ the wormhole geometry is free of singularities.

An interesting property about wormholes that is usually analyzed is its traversability. It is determined by the flare-out condition \cite{MT}.
This condition states that
\begin{eqnarray}
\label{flareout}
\Omega'(r) < \frac{\Omega(r)}{r} &,& \Omega(r) - r<0.
\end{eqnarray}
The shape function satisfies the flare-out condition (\ref{flareout}) provided that
\begin{equation} \label{flare}
\frac{3-2 \lambda}{1- 2\lambda} < 1,
\end{equation}
which is valid for all $\lambda$ such that $\lambda>\frac{1}{2}$. It is worthwhile to mention that this is the same interval wherein the metric is asymptotically flat.



\section{Energy Conditions} \label{sec3}

\indent\indent The study of the energy conditions of wormhole solutions is of paramount importance, since the usual solutions of this nature usually violate the null energy condition, Ref.\cite{MT}, and thus, these solutions are generated by some kind of exotic matter. In this section, we examine the validity of the energy conditions given by Ref. \cite{wald} upon the modified wormhole solution with isotropic pressures (\ref{solucao2}).

Let us start with the Null energy condition (NEC), which states that
$\rho + p \geq 0$.
From the energy density (\ref{energydensity}) and the eos (\ref{eos}), we obtain
\begin{equation}
\bigg(  \frac{2-\lambda}{1-2\lambda}   \bigg)\frac{r^{\frac{4 \lambda}{1- 2 \lambda}}}{k r_{0} \ ^{\frac{2}{1- 2 \lambda}}} \geqslant 0.
\end{equation}
Thus, the NEC is satisfied provided the Lorentz violating parameter $\lambda$ belongs to the intervals $-\infty<\lambda<\frac{1}{2}$ or $2<\lambda<\infty$.

\begin{figure}[h] 
\centering 
\includegraphics[scale=0.3]{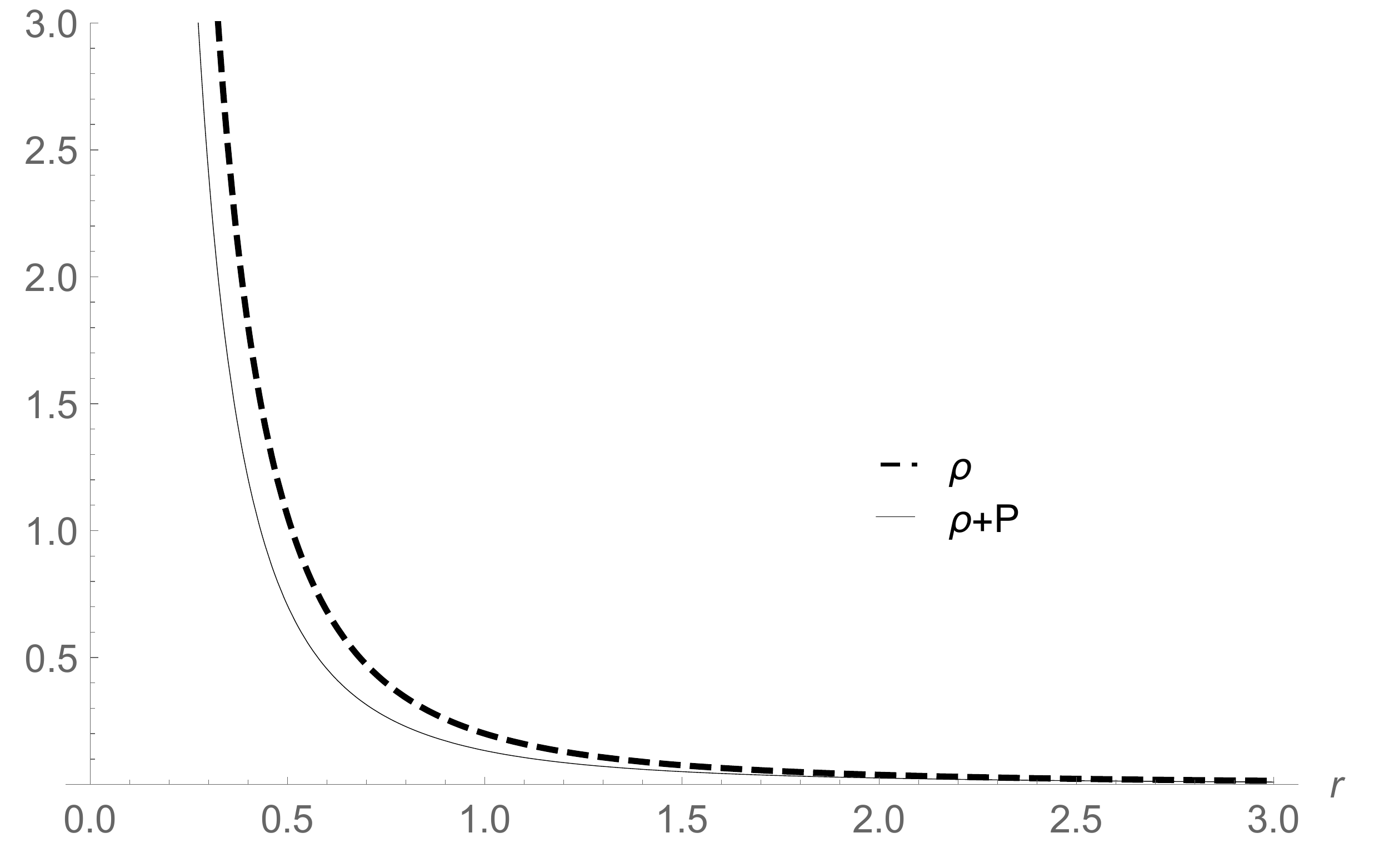} 
\caption{Energy density $\rho$ (dashed line) and $\rho + P$ (thin line) for $\lambda=3$.}
\label{fig1}
\end{figure}


The Weak energy condition (WEC) states that $\rho + p \geqslant 0$ and $\rho \geqslant 0$.
From the Eq.(\ref{energydensity}) the energy density is non-negative in the intervals $\lambda<\frac{1}{2}$ and $\lambda>2$. In the figure (\ref{fig1}) we plotted the energy density (dashed line) and $\rho + P$ (thin line) for $\lambda=3$. The figure portraits a localized source with negative pressure.


%

The Strong energy solution (SEC) holds provided that $\rho + p \geqslant 0$ and $\rho + 3p \geqslant 0$.
Since the source satisfies the equation of state (eos) $P=-\frac{1}{3}\rho$, then $\rho+3P=0$, regardless the value of $\lambda$. Therefore, for $\lambda>2$ the NEC, WEC, SEC and the flare-out condition are all satisfied.


\section{Light Deflection by Gauss-Bonnet Theorem} \label{sec4}
In this section we analyze the light deflection by static spherically symmetric wormhole of gravity nonminimally coupled to a VEV of the Kalb-Ramond field. Since we are dealing in the weak field limit, than it is convenient to use the Gauss-Bonnet theorem \cite{Gibbons, Kimet}. Considering an oriented surface,  let us define a domain $ (D, \chi, g) $ being a subset of a compact manifold with a boundary $(\mathcal{D}_{R}, \chi, g)$ with Euler characteristic factor $ \chi $ and metric $ g $ \cite{Gibbons}. So the Gauss-Bonnet theorem, which gives us the deflection angle, can be understood as
\begin{equation}
\int \int_{\mathcal{D}_{R}} K dS + \oint_{\partial \mathcal{D}_{R}} k dt + \sum_{i} \theta_{i} = 2 \pi \chi (\mathcal{D}_{R}),
\label{z222}
\end{equation}
where $ \mathcal{D}_{R} $ is a region that contains the source of the light rays, the observer's referential and the center of the lens. This region being limited by the outermost rays, represented by the boundary $ \partial \mathcal {D}_{R} $. The term $ k $ is identified as a geodesic curvature. 
For simplicity, we can consider that for the observer the sum of the exterior angles $ \theta_{i} $ becomes $\pi $, as long as the radial distance is $ R \rightarrow \infty $ \cite{Suzuki}. The characteristic factor of Euler $ \chi (\mathcal {D}_{R}) $ is for our purposes made equal to $ 1 $. 
Then, the equation \eqref{z222} becomes
\begin{equation}
\int \int_{\mathcal{D}_{R}} K dS + \oint_{\partial \mathcal{D}_{R}} \kappa \,dt = \pi,
\label{x79}
\end{equation}
where the term $ dS $ is the surface element.

Now considering our metric in general form,
\begin{equation}
ds^2 = g_{tt}\, dt^2 -  g_{rr} \, dr^2 - r^2\, d\theta^2 - r^2 sin^2\theta \, d\phi^2,
\label{a55}
\end{equation}
for wormhole with the configuration $g_{tt} = 1$ and $g_{rr}  =\frac{1 }{ 1 -  \left(\frac{r}{r_{0}}\right)^{\frac{2}{1 - 2\lambda}}}$. The optical path can be defined using the metric \eqref{a55} using the condition of null geodesic
$ds^2 = 0$. Then the optical path metric is written as
\begin{equation}
dt^2 = \bar{g}_{ij} dx^{i} dx^{j} = \bar{g}_{rr} dr^2 + \bar{g}_{\phi\phi} d\phi^2,
\label{zz00}
\end{equation}
where we consider that the photon moving in the equatorial plane $\theta = \frac{\pi}{2}$. 
Based on the \eqref{zz00} equation, the optical path metric is explicitly expressed as
\begin{equation}
 dt^2 = \frac{dr^2 }{ 1 -  \left(\frac{r}{r_{0}}\right)^{\frac{2}{1 - 2\lambda}} }  + r^2 d\phi^2.
 \label{x77}
\end{equation}
Optical path terms are identified as $ \bar{g}_{rr} = \frac{1 }{ 1 - \left( \frac{r}{r_{0}}\right)^{\frac{2}{1 - 2\lambda}}} $ and $\bar{g}_{\phi\phi} = r^2$.

The geodesic curvature is defined by \cite{Ju-Ov}:
\begin{equation}
\kappa(C_{R}) = \mid\bigtriangledown_{\dot{C}_{R}} \dot{C}_{R}\mid,
\label{x92}
\end{equation}
taking the radial component of the equation \eqref{x92},
\begin{equation}
\left( \bigtriangledown_{\dot{C}_{R}} \dot{C}_{R}\right)^{r} = \dot{C}_{R}^{\phi} \left(\partial_{\phi} \dot{C}_{R}^{r}\right)  + \Gamma_{\phi\phi}^{r} \left(\dot{C}_{R}^{\phi} \right)^2,
\label{x101}
\end{equation}
where  $\Gamma_{\phi\phi}^{r}$ is the Christoffel symbol referring to the optical path of the equation \eqref{x77}, which results in
\begin{equation}
 \Gamma_{\phi\phi}^{r} = \frac{1}{2} \,\bar{g}^{rr} \frac{\partial \bar{g}_{\phi\phi}}{\partial \, r} = - r\, \left(   1 -  \left(\frac{r}{r_{0}}\right)^{\frac{2}{1 - 2\lambda}}\right) .
\end{equation}
The radial distance can be considered a constant so large that it tends to infinity, $ r \equiv R \rightarrow \infty$. Since $ \dot{C}_{R} $ does not depend on $ \phi $, then only the last term of the equation \eqref{x101} contributes as well as the term $\dot{C}_{R} = 1/R $ \cite{Sakalli}. Therefore, the equation \eqref{x92} tells us that
\begin{eqnarray}
lim_{R \rightarrow \infty} \, k\left( C_{R}\right)  = & lim_{R \rightarrow \infty} \vert \bigtriangledown_{\dot{C}_{R}} \dot{C}_{R} \vert = \frac{\left(  1 -  \left(\frac{R}{r_{0}}\right)^{\frac{2}{1 - 2\lambda}} \right)  }{ R} \nonumber\\ & \rightarrow \frac{1}{ R}.
\label{x200}
\end{eqnarray}
The last step in equation \eqref{x200} is valid for every $\lambda$ that verifies the relationship $\frac{2}{1 - 2\lambda} < 0$. 

Now the Gauss curvature $ K $ can be calculated by the expression found in Refs.\cite{Ara, Gun}. In the case of $ g_{tt} (r) = 1 $ and taking into account that the $\bar{g}_{rr}$ e $\bar{g}_{\phi\phi}$ are independent of $\phi$, 
the equation \label{eq: GB2} can takes the form
\begin{equation}
  K = \frac{1}{\bar{g}_{rr}}\, \frac{1}{r}\,\frac{d\, ln\left( \sqrt{\bar{g}_{rr}} \right) }{d\, r}.
\label{qq1}
\end{equation}
The infinitesimal surface element is defined by 
\begin{equation}
  dS = \sqrt{\bar{g}_{rr}} \,r\, dr\, d\phi,
  \label{qq2}
  \end{equation}
where the term
  \begin{equation}
  \bar{g}_{rr}  = \frac{1}{ 1 - \left(\frac{r}{r_{0}}\right) ^{\frac{2}{1 - 2\lambda}} }.
    \label{z00}
    \end{equation}
 Substituting the equation \eqref{qq1} in the equation \eqref{qq2} we obtain
  \begin{equation}
   K\, dS = \frac{1}{ \sqrt{\bar{g}_{rr}}}\frac{d\, ln \left(\sqrt{\bar{g}_{rr}} \right) }{d\, r}\, dr\,d\phi = - \frac{d}{dr} \,\left( \frac{1}{\sqrt{\bar{g}_{rr}}}\right)  \, dr\, d\phi.
   \label{x201}
   \end{equation}
Remembering that $ R = const $, from the equation \eqref{x77} we obtain, 
\begin{equation}
 dt = R\, d\phi.
 \label{x100}
 \end{equation} 
 Replacing the equations \eqref{x200} and \eqref{x100} in the equation \eqref{x79} we simplify our calculations and thus obtain
 \begin{equation}
\int \int_{\mathcal{D}_{\infty}} K dS + \int_{0}^{ \pi +\alpha} \frac{1}{ R}\, R\,d\phi = \pi.
\label{x166}
\end{equation}
Since the rays of light come from a source at infinity up to such radial distance and remembering that we are in the weak field regime, the rays are approximately straight. So we can  use a condition $r = \frac{\sigma}{ sin(\phi)}$ \cite{Og}, where $\sigma$ is the impact parameter. If we substitute the equations \eqref{z00} and \eqref{x201} in the equation \eqref{x166} we find 
\begin{eqnarray}
\alpha =  & -\int_{0}^{\pi} \int_{\frac{\sigma}{ sin(\phi)}}^{\infty} K dS  = \nonumber \\
&\int_{0}^{\pi} \int_{\frac{\sigma}{ sin(\phi)}}^{\infty} \frac{d}{dr}\, \sqrt{1 - \left(\frac{r}{r_{0}}\right)^{\frac{2}{1 - 2\lambda}}} \, dr\, d\phi,
\label{x167}
\end{eqnarray}
and integrating it with respect to the variable $ r $ we obtain, 
\begin{eqnarray}
\alpha =  &\int_{0}^{\pi} \sqrt{1 -\left(\frac{r}{r_{0}}\right) ^{\frac{2}{1 - 2\lambda}}}  \Big{\vert}_{\frac{\sigma}{Sin(\phi)}}^{\infty} d\,\phi \Rightarrow \alpha = \nonumber \\
&\int_{0}^{\pi} \left(1 - \sqrt{1 - \left( \frac{\sigma}{r_{0}\, sin \phi}\right)^{\frac{2}{1 - 2\lambda}}}\right)\, d \phi .
\label{x2099}
\end{eqnarray}

Despite the complexity of 
Eq. \eqref{x2099}, analytical expressions can be found for  $\lambda = \frac{3}{2}$ and $\lambda = 1$. For the value $ \lambda = \frac{3}{2} $, the deflection angle takes the form
\begin{equation}
\alpha = \pi - \frac{4 \sqrt{\sigma - r_{0}} \sqrt{\sigma}\, E\left(\frac{\pi }{4}, -\frac{2\, r_{0}}{\sigma-r_{0}}\right)}{\sigma},
\label{x999}
\end{equation} 
where $ E(x, y) $ refers to the elliptical function of the second type. Since we are in weak deflection limit approximation and write the new variable $ z = \frac{r_0}{\sigma} $, we can expand the equation \eqref{x999} in terms of z, and we find that
\begin{equation}
   \alpha \simeq \frac{r_{0}}{\sigma} +\pi \left(\frac{r_{0}}{4\,\sigma} \right)^2 +\mathcal{O}\left(\frac{r_0}{\sigma}\right)^3.
   \label{x1111}
\end{equation}

\begin{figure}[h]
\centering
\includegraphics[scale=0.5]{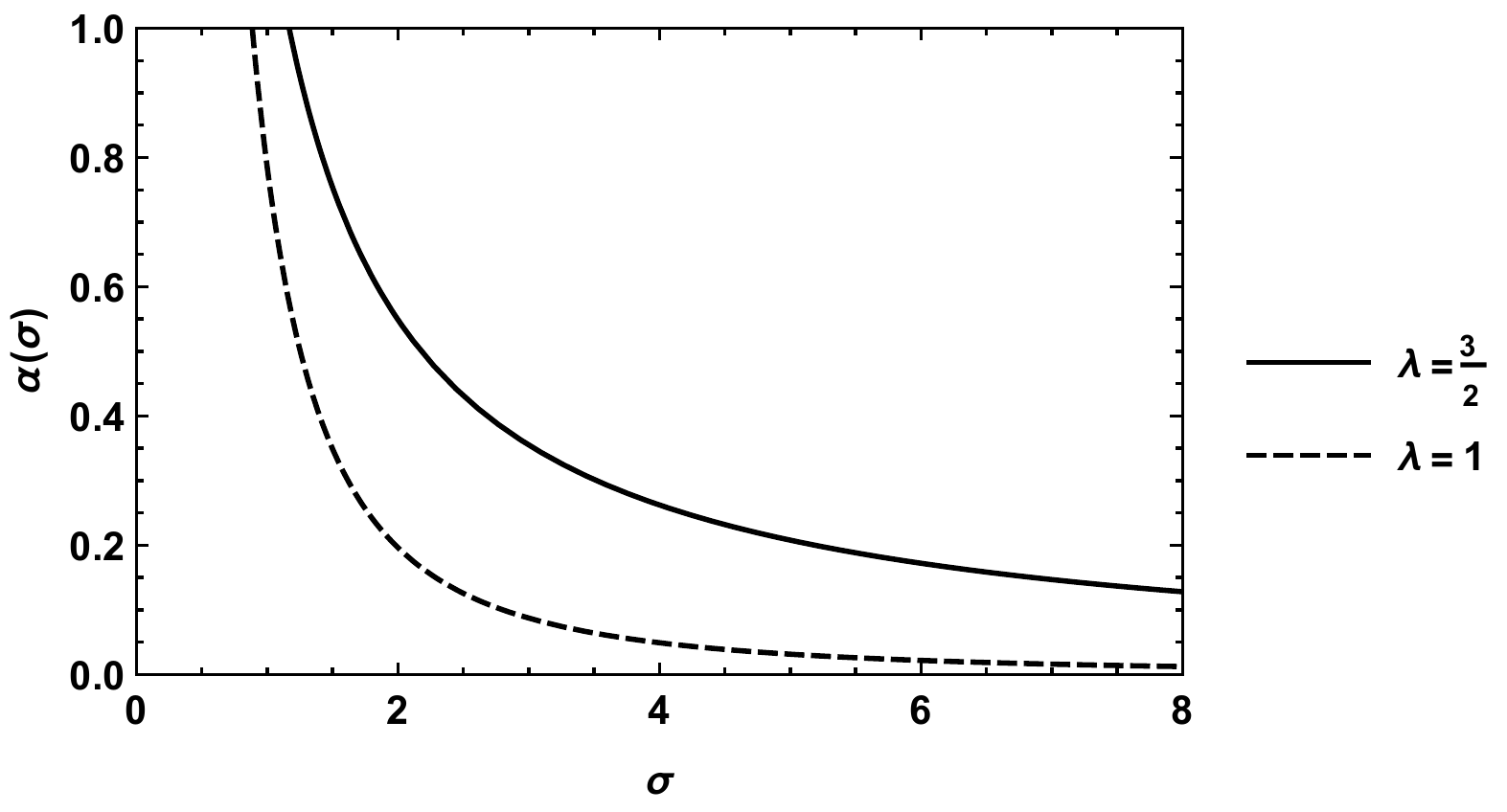}
\caption{Light deflection angle as a function of the impact parameter $\sigma$ for the values $\lambda = 1$ and $\lambda = \frac{3}{2}$, with $r_{0} = 1$ }
\label{zaxp}
\end{figure}

In the second case, for the value $ \lambda = 1 $, the last term of the equation \eqref{x2099} tells us that the angle of deflection is
\begin{equation}
\alpha = \pi - 2 E\left( \frac{r_{0}^2}{\sigma^2}\right).
\label{x1000}
\end{equation}
Expanding the equation \eqref{x1000} in terms of  $ z = \frac{r_0}{\sigma} $ we find
\begin{equation}
    \alpha \simeq \pi \left(\frac{r_{0}}{2\, \sigma}\right)^2 +\mathcal{O}\left(\frac{r_0}{\sigma}\right)^3.
    \label{x1112}
\end{equation}
It is worthwhile to compare the light deflection angle due to KR VEV with other solutions. Unlike the Bumblebee model, the KR correction \eqref{x1111} exhibits a $\frac{r_{0}}{\sigma}$ term for $\lambda= \frac{3}{2}$. Such $\frac{r_{0}}{\sigma}$ term arises in rotating wormhole solutions, as in Ref.\cite{Ka}. For $\lambda = 1$, the angle of deflection has the same form of the Ellis wormhole \cite{Kimet}, and the Lorentz violating corrections only arise at third-order in $\frac{r_{0}}{\sigma}$.
The figure (\ref{zaxp}) depicts the behaviour of the deflection angle $\alpha$ with respect to the impact parameter $\sigma$ for $\lambda=1$ and $\lambda=3/2$.
Likewise the bumblebee wormhole \cite{Jusu}, the KR VEV introduces a Lorentz violating dependence upon the deflection of light.

\section{Final remarks and perspectives}
\label{sec5}
We obtained a static and spherically symmetric wormhole solution of gravity nonminimally coupled to a Kalb-Ramond VEV field. The self-interaction potential breaks the KR gauge invariance and produces a VEV background tensor field which violates the local Lorentz symmetry.

The vacuum background tensor was chosen in order to vanish the self-interaction potential and the vacuum hamiltonian. Further, the VEV is perpendicular to the timelike and spacelike Killing vectors and then, the Lorentz violation preserves the static and spheric symmetries of the gravitational vacuum. By assuming a non-minimal coupling between the KR VEV and the Ricci tensor, a zero-tidal-force wormhole by a perfect fluid was found, for which the condition $p_{r} = p_{\theta}$ must be required. For traversable wormhole solution, the LV parameter must be $\lambda > 1/2$. Moreover, the matter source has a negative pressure with equation of state $p=-\rho/3$.

The analysis the energy conditions of the LV source revealed that for the Lorentz symmetry breaking parameter in the range $\lambda > 2$, the the null, weak and strong energy conditions hold. In addition, the flare-out condition is also satisfied in this range.

By employing the Gauss-Bonnet theorem, we studied the effects of the LSB wormhole on the light deflection. The behaviour of the deflected angle $\alpha$ with respect to the impact parameter is similar to one found in the Bumblebee LBS \cite{Jusu} and in rotating wormhole \cite{Ka}, in the case $ \lambda = \frac{3}{2}$. In the case of $\lambda = 1$ the behaviour of the deflected angle is the same of the Ellis wormhole \cite{Kimet}. Yet, the KR LSB increases the rate that deflection angle decreases.


As possible extensions of the present work, we point out the stability analysis of the solution found by considering the corrections due to fluctuations of the KR field around the VEV. The analysis for magnetic-monopole and dyon-like KR VEV configurations are also in order. 
The effects of a coupling between the KR field and the Riemann tensor is another important development.

\section*{Acknowledgments}
\hspace{0.5cm}The authors thank the Conselho Nacional de Desenvolvimento Cient\'{\i}fico e Tecnol\'{o}gico (CNPq), grants n$\textsuperscript{\underline{\scriptsize o}}$ 312356/2017-0 (JEGS) and n$\textsuperscript{\underline{\scriptsize o}}$ 308638/2015-8 (CASA) for financial support.

\end{document}